\documentclass[useAMS]{mn2e}
\usepackage{times}
\input{epsf}
\usepackage{graphicx}
\usepackage{lscape}

\def\mnras{MNRAS}
\def\apj{ApJ}
\def\apjl{ApJL}
\def\aap{A \& A}
\def\araa{ARAA}
\def\pasp{PASP}
\def\pasj{PASJ}
\def\apjs{ApJS}
\def\aaps{A \& A S}

\title[Analysing the atolls]
{Analysing the atolls: X-ray spectral transitions of accreting neutron
stars}

\author[J. C. Gladstone, C. Done, M.~Gierli\'nski]
{Jeanette Gladstone$^1$$^{*}$, Chris Done$^1$ and  Marek~Gierli\'nski$^{1,2}$\\
$^1$Department of Physics, University of Durham, South Road, Durham DH1 3LE,
UK\\
$^2$Obserwatorium Astronomiczne Uniwersytetu Jagiello{\'n}skiego, 30-244
Krak{\'o}w, Orla 171, Poland\\
$^{*}$j.c.gladstone@durham.ac.uk}

\date{Submitted to MNRAS}
\pagerange{\pageref{firstpage}--\pageref{lastpage}}
\pubyear{2005}

\begin{document}

\topmargin = -0.5cm

\maketitle
\label{firstpage}

\begin{abstract}

We systematically analyze all the available X-ray spectra of disc
accreting neutron stars (atolls and millisecond pulsars) from the
{\it RXTE} database. We show that while these all have similar
spectral evolution as a function of mass accretion rate, there are
also subtle differences. There are two different types of hard/soft
transition, those where the spectrum softens at all energies,
leading to a diagonal track on a colour-colour diagram, and those
where only the higher energy spectrum softens, giving a vertical
track. The luminosity at which the transition occurs is {\em
correlated} with this spectral behaviour, with the vertical
transition at $L/L_{\rm Edd}\sim 0.02$ while the diagonal one is at
$\sim 0.1$. Superimposed on this is the well known hysteresis
effect, but we show that classic, large scale hysteresis occurs only
in the outbursting sources, indicating that its origin is in the
dramatic rate of change of mass accretion rate during the disc
instability. We show that the long term mass accretion rate
correlates with the transition behaviour, and speculate that this is
due to the magnetic field being able to emerge from the neutron star
surface for low average mass accretion rates. While this is not
strong enough to collimate the flow except in the millisecond
pulsars, its presence may affect the inner accretion flow by
changing the properties of the jet.

\end{abstract}


\section{Introduction}

Neutron stars represent the most dense form of matter known before
black hole formation. The equation of state of such material is not
well understood but typical expected size scales of $R\sim 10$~km
for a $1.4M_\odot$ object give $M/R$ similar to that of the last
stable orbit around a Schwarzschild black hole. The gravitational
potential seen by the accreting material is then very similar in
neutron stars and black holes, but with the obvious difference that
the material may then fall seamlessly and invisibly from the edge of
the disc through the event horizon in the black holes (but see e.g.
Krolik, Hawley \& Hirose 2005 for an alternative view), but must
impact onto the solid surface for the neutron stars. This boundary
layer emission should give a clear distinction between the two types
of systems, yet in practice it can be quite hard to discriminate
between the disc accreting, low magnetic field neutron stars and
Galactic black holes (GBH). Both show hard spectra at low mass
accretion rates (e.g. Barret \& Vedrenne 1994; Barret et al. 2000)
while at higher mass accretion rates the black holes can show low
temperature, optically thick Comptonized emission from a corona
which can look similar to an optically thick boundary layer in the
neutron stars systems (the black hole `very high state', Done \&
Gierli{\'n}ski 2003, hereafter DG03).

Nonetheless, DG03 showed that there are clear differences between
the black hole and neutron star systems. They used the huge database
now available on these objects from the {\it Rossi X-ray Timing
Explorer} ({\it RXTE}) to build a picture of the X-ray spectral
evolution as a function of overall mass accretion rate (as measured
by luminosity, parameterized in units of the Eddington rate,
$L/L_{\rm Edd}$).  They used data from GBH and both subclasses of
disc accreting, low magnetic field neutron stars namely atolls and Z
sources (see e.g. Hasinger \& van der Klis 1989). The atolls and GBH
cover the same luminosity range in $L/L_{\rm Edd}$, from $\sim
10^{-3}$ to $\sim 0.5-1$, so their accretion flows should be
directly comparable (Z sources are typically more luminous, and may
have residual magnetic fields: Hasinger \& van der Klis 1989). DG03
showed that the GBH as a class were all consistent with the same
spectral evolution as a function of mass accretion rate, as were the
atolls, but that this behaviour {\em differed substantially} between
the two types of system even though they covered the same range in
$L/L_{\rm Edd}$.

DG03 interpreted this in the light of the recent advances in
accretion flow models, which suggest that at low mass accretion
rates the optically thick, cool, inner disc can evaporate into an
optically thin, hot flow (e.g.  an advection dominated accretion
flow: Narayan \& Yi 1996 or a jet dominated accretion flow: Falcke,
Kording \& Markoff 2004). These models are currently controversial
as they conflict with the occasional detection of extremely broad
iron lines in GBH spectra in this state (Miller et al 2002;
Miniutti, Fabian \& Miller 2004; Miller et al 2006), though these
extreme widths could also be an artifact of complex absorption (Done
\& Gierli{\'n}ski 2006). Nonetheless, the truncated disc/hot inner
flow models are very attractive as they can qualitatively explain a
wide range of observed properties in X-ray binaries.  This includes
the spectral evolution of the GBH, which can be broadly explained by
the transition radius between the disc and hot flow progressively
decreasing as the mass accretion rate increases, while the atolls
are consistent with an identical accretion flow behaviour but with
the addition of the expected boundary layer/surface emission (DG03).

These observational differences between GBH and atolls are consistent
with the existence of the event horizon, the most fundamental
predicted property of a black hole (see also McClintock, Narayan \&
Rybicki 2004; Narayan \& Heyl 2002; Garcia et al. 2001; Sunyaev \&
Revnivtsev 2000).  However, the sample of atolls used by DG03 was
fairly small, with only 4 objects (1 of which was an accreting
millisecond pulsar). This contrasts with their GBH selection, which
included all available objects which were not heavily absorbed. Here
we present a much wider study using all the available atolls in
the {\it RXTE} database which are not heavily absorbed to investigate
their behaviour in more detail. We find that while all the atolls (and
millisecond pulsars) broadly show the same spectral evolution, there
are some subtle differences. Some of these can be connected to
spin/inclination effects, but there are also some differences in
transition behaviour which correlate with long term mass accretion
rate. We speculate that these may be linked through the jet being
affected by any remnant polar magnetic field, which in turn depends on
the accretion rate.

\section{Sample Selection and Data Analysis}

We searched the {\it RXTE} database for low magnetic field, disc
accreting neutron star systems. We do not include Z sources (Sco
X-1, GX349+2, GX17+2, Cyg X-2, GX5--1, GX340+0 e.g. Kuulkers et al
1996, Cir X-1, LMC X-2, Hasinger \& van der Klis 1989) as we were interested 
in the range $10^{-3}<L/L_{Edd}<1$ to compare with the galactic black holes 
from DG03. We also exclude all objects with heavy absorption, defined as $N_H > 3\times
10^{22}$ cm$^{-2}$, as above this the spectral decomposition becomes
more uncertain. Given that most systems are in the Galactic plane,
this removes many objects (4U 1624--49, GRO J1744--28, GRS
1747--312, SAX J1747.0--2853). The absorption criteria also excludes
objects where there is substantial intrinsic absorption. This can be
continuous, completely blocking our view of the continuum source so
that the only scattered flux is seen (the accretion disc corona
sources), or the absorption can be episodic, giving discrete dips in
the lightcurve (the dippers). We exclude all accretion disc corona
sources (2S 0921-630, 4U 1624-49, 4U 1822-371, MS 1603+206), and
dippers where the absorption events are so frequent that most
observations are affected by them (EXO~0748-676), but we do include
dippers where the absorption is limited to contaminating only a few
observations (which we remove, see next section).  Details of all
included sources are shown in Table \ref{tab:sources}.

\begin{table*}
\leavevmode
\begin{center}
\begin{tabular}{cccccccccc}
\hline
Source Name & Alternate Name & $N_H$ & Distance & Persistent & PCA State & Dips & Period & Spin  & References\\
  &  &  &  & or & or &  &  &  & \\
  &  & ($10^{22}$ cm$^{-2}$) & (kpc) & Transient & Transition &  & (hrs) & (Hz) & \\
\hline
4U 0614+091      &               & 0.28     & 3     & P  & T &   & 0.25 &     & 1,2,4 \\
4U 0919--54      & 2S 0918--549  & 0.32     & 4.9   & P  & T &   & 0.42 &     & 4,6,7 \\
4U 1254--690     & GR Mus        & 0.291    & 13    & P  & B & Y & 3.9  &     & 1,3,8,9,10,11 \\
4U 1543--624     &               & 0.27     & 7     & P  & B &   & 0.3  &     & 3,4,7,12,13,14 \\
4U 1556--605     &               & 0.296    & 4     & P  & B &   & 9.1  &     & 1,3 \\
4U 1608--52      & GX 331--01    & 1.5      & 3.6   & ST & T &   & 288  & 619 & 1,3,15,16,17 \\
4U 1626--67      &               & 0.069    & 8     & P  & I &   & 0.68 &     & 18,19,20 \\
4U 1636--53      &               & 0.37     & 5.9   & P  & T &   & 3.9  & 582 & 1,15,21 \\
4U 1702--43      & Ara X-1       & 1.2      & 7.3   & P  & T &   &      & 330 & 1,3,5,8\\
4U 1704--30      & MXB 1659--298 & 0.35     & 12    & LT & T & Y & 4.2  & 567 & 22,23,24 \\
4U 1705--44      &               & 1.47     & 7.4   & P  & T &   &      &     & 1,3,25 \\
4U 1708--40      &               & 2.9      & 8     & P  & B &   &      &     & 26 \\
4U 1711--35      & 2S 1711--339  & 1.5      & 8     & ST & T &   &      &     & 1,27 \\
4U 1724--307     &               & 1        & 6.6   & P  & T &   &      &     & 1,28,29 \\
4U 1728--16      & GX 9+9        & 0.232    & 5     & P  & B &   &      &     & 1,2,3,30 \\
4U 1728--34      & GX 345--0     & 2.5      & 4.75  & P  & T &   &      & 363 & 1,15,31,32,33 \\
4U 1730--33      & Rapid Burster & 1.6      & 8     & ST & B &   &      &     & 1,3,34 \\
4U 1735--44      &               & 0.46     & 9.2   & P  & B &   & 4.65 &     & 1,2,3,5,35 \\
4U 1744--26      & GX3+1         & 1.76     & 8.5   & P  & B &   &      &     & 1,2,3,5 \\
4U 1746--37      &               & 0.295    & 10.7  & P  & T & Y & 5.16 &     & 1,3,8,36,37 \\
4U 1758--20      & GX 9+1        & 1.433    & 8.5   & P  & B &   &      &     & 1,2,3,5 \\
4U 1812--12      &               & 1.6      & 4     & P  & I &   &      &     & 1,38 \\
4U 1820--303     &               & 0.24     & 5.8   & P  & T &   & 0.19 &     & 1,3,39 \\
4U 1837+04       & Ser X-1       & 0.51     & 8.4   & P  & B &   &      &     & 1,3,5 \\
4U 1850--08      &               & 0.39     & 6.8   & P  & T &   & 0.35 &     & 3,40,41 \\
4U 1908+005      & Aql X-1       & 0.96     & 2.5   & ST & T &   & 19   & 549 & 1,3,15,17,42 \\
4U 1915--05      & 2S 1912--05   & 0.2      & 9.3   & P  & T & Y & 0.8  & 272 & 1,3,8,43,44,45,46 \\
GS 1826--236     &               & 2.4      & 6     & P  & I &   & 2.1  & 611 & 1,47,48,49 \\
KS 1731--26      &               & 1        & 7     & LT & T &   &      & 524 & 1,15,47,50,51 \\
SLX 1732--304    &               & 1.2      & 15    & LT $^{*}$ & B &   &      &     & 52,53 \\
SLX 1735--269    &               & 1.47     & 8.5   & P  & T &   &      &     & 1,48,54,55 \\
XTE J1709--267   &               & 0.44     & 8.8   & ST & T &   &      &     & 56 \\
XTE J1806--246   &               & 0.5      & 8     & ST $^{*}$ & T &   &      &     & 1,57,58 \\
XTE J2123--058   &               & 0.66     & 8     & ST & T &   &      &     & 1,59 \\
\hline
IGR J00291+5934   &                  & 2.8   & 4    & ST & I &  & 2.46 & 599 & 60,61,62 \\
XTE J0929--314    &                  & 0.21  & 10   & ST & I &  & 0.73 & 185 & 61,63,65 \\
XTE J1751--305    &                  & 1.01  & 8    & ST & I &  & 0.7  & 435 & 61,63,64,65 \\
XTE J1807--294    &                  & 0.5   & 8    & ST & I &  & 0.67 & 190 & 61,63\\
XTE J1808--369    & SAX J1808.4--3658 & 0.122 & 3.15 & ST & I &  & 2    & 401 & 17,61,63,66 \\
XTE J1814--338    &                  & 0.167 & 8    & ST & I &  & 4.27 & 314 & 61,67,68 \\
\end{tabular}
\end{center}
\caption{A list of atoll (above the dividing line) and millisecond pulsar (below the line) sources with observations available from the
{\it RXTE} database. Letters in column 4 refer to: P - persistent, ST - transient (can be recurrent), LT - long term transients (outburst $>3$ month), found using the RXTE Galactic Centre Observations (http://lheawww.gsfc.nasa.gov/users/craigm/galscan/main.html) and the RXTE ASM Weather Map definitive lightcurves (http://heasarc.gsfc.nasa.gov/xte\_weather). Classifications marked $^{*}$ denote sources for which long term lightcurves were not available. In this case classifications were obtained via refereed articles. Letters in column 5 refer to: B - banana (soft) state only, I - island (hard) state only, T - transitions between both states. References are:
$^{1}$Liu et al. 2001,
$^{2}$Schultz 1999,
$^{3}$Christian \& Swank 1997,
$^{4}$Nelemans et al. 2004,
$^{5}$Fender \& Hendry 2000,
$^{6}$Jonker et al. 2001,
$^{7}$Juett \& Chakrabarty 2003,
$^{8}$Diaz Trigo et al. 2006,
$^{9}$Smale et al. 2002,
$^{10}$Courvoisier et al. 1986,
$^{11}$in 't Zand et al. 2003,
$^{12}$Shultz 2002,
$^{13}$Wang \& Chakrabarty 2004,
$^{14}$Farinelli et al. 2003,
$^{15}$Piro \& Bilsten 2005,
$^{16}$Gierli{\'n}ski \& Done 2002,
$^{17}$Garcia et al. 2001,
$^{18}$Owens et al. 1997,
$^{19}$Chakrabarty 1997,
$^{20}$Angelini et al. 1995,
$^{21}$Wijnands 2001,
$^{22}$Oosterbroek et al. 2001,
$^{23}$Wijnands et al. 2001,
$^{24}$Oosterbroek et al. 2001,
$^{25}$Di Salvo et al. 2005,
$^{26}$Migliari et al. 2002,
$^{27}$Wilson et al. 2003,
$^{28}$Molkov et al. 2000,
$^{29}$Emelyanov et al. 2002,
$^{30}$Yao \& Wang 2005,
$^{31}$Di Salvo et al. 2000,
$^{32}$Shaposhnikov et al. 2003,
$^{33}$Migliari et al. 2003,
$^{34}$Falanga et al. 2004,
$^{35}$Cornelisse et al. 2000,
$^{36}$Jonker et al. 1999,
$^{37}$Ba{\l}uci{\'n}ska-Church et al. 2003,
$^{38}$Barret et al. 2003,
$^{39}$Shaposhnikov \& Titarchuk 2004,
$^{40}$Sidoli et al. 2005,
$^{41}$Homer et al. 1996,
$^{42}$Chevalier et al. 1999,
$^{43}$Blosser et al. 2000,
$^{44}$Chevalier et al. 1999,
$^{45}$Maccarone \& Coppi2002
$^{46}$Galloway et al. 2001,
$^{47}$Barret et al. 1999,
$^{48}$Galloway at al. 2003,
$^{49}$Thompson et al. 2005,
$^{50}$Mignani et al. 2002,
$^{51}$Burderi et al. 2002,
$^{52}$Pavlinsky et al. 2001,
$^{69}$Cackett et al. 2006,
$^{54}$Molkov et al. 2005,
$^{55}$Wijnands \& van der Klis 1999(a),
$^{56}$Jonker et al. 2004,
$^{57}$Wijnands \& van de Klis 1999(b),
$^{58}$Revnivtsev et al. 1999,
$^{59}$Tomsik et al. 2004,
$^{60}$Galloway et al. 2005,
$^{61}$Poutanen 2005,
$^{62}$Jonker et al. 2005,
$^{63}$Campana et al. 2005,
$^{64}$Campana et al. 2003,
$^{65}$Gierli{\'n}ski \& Poutanen 2005,
$^{66}$Wijnands 2003,
$^{67}$Krauss et al. 2005,
$^{68}$Strohmayer et al. 2003.}

\label{tab:sources}
\end{table*}

\section{Spectral evolution of atolls and millisecond pulsars}

The standard products spectra (and associated background and response
files) from the Proportional Counter Array (PCA), available from the
High Energy Astrophysics Science Archive Research Centre (HEASARC)
database were gathered for each source, giving a total of over 3000
spectra. Following DG03 we fit these with a model of the disc emission
and a Comptonized continuum, described by {\sc diskbb} (Mitsuda et
al. 1984) and {\sc thcomp} (Zdziarski et al. 1996) in {\sc xspec},
together with a smeared edge and gaussian line to approximately mimic
the reflected spectrum. The absorption was fixed at the Galactic value
for each object (see Table \ref{tab:sources}). This model gave a good
fit to all the spectra ($\chi^{2}_{\nu}<1.5$).

Following DG03, the best-fitting model to each spectrum was
integrated to form intrinsic soft and hard colours (i.e. absorption
corrected and independent of the instrument response), defined as
flux ratios in the bandpasses 4--6.4/3--4 and 9.7--16/6.4--9.7~keV,
respectively. The unabsorbed model was also used to estimate the
bolometric flux in the 0.01--1000~keV bandpass, and this converted
into a luminosity using the distance estimate given in Table
\ref{tab:sources}. As discussed by DG03, the colours are fairly
robust to changes in the model spectra as long as the source is not
heavily absorbed and the advantage of using intrinsic colours (as
opposed to the more widely used instrument colours) is that many
objects can be directly compared on the same plot. The bolometric
fluxes are more uncertain, as the model is extrapolated outside of
the observed energy bandpass. Nonetheless, this model corresponds to
expected physical components, and we fix a lower limit to the disc
temperature of $\ge 0.4$~keV and an upper limit to the Comptonizing
temperature of $\le 100$~keV so the continuum components cannot
produce arbitrarily large luminosities in the unobserved energy
ranges (DG03).

\begin{figure*}
\leavevmode
\begin{center}
\epsfxsize=14cm \epsfbox{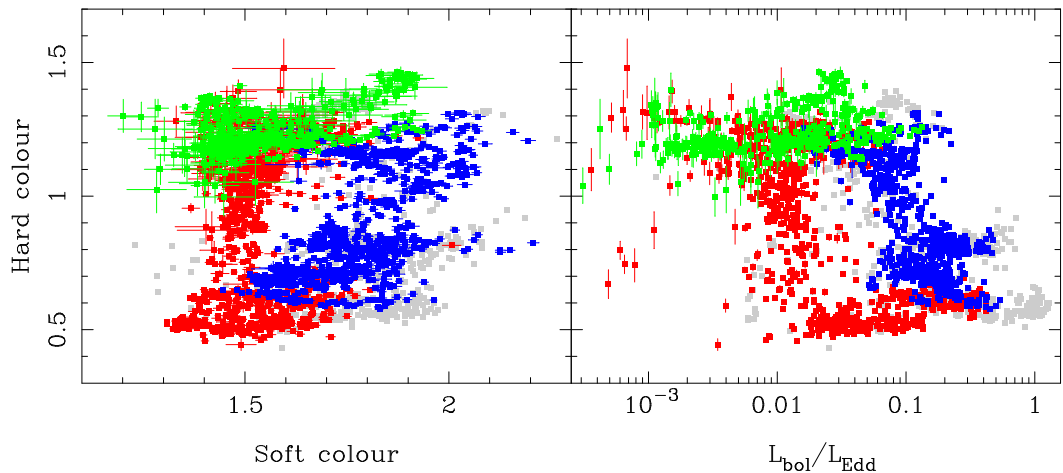}
\end{center}
\caption{Combined colour-colour colour-luminosity plots for the {\it
RXTE} PCA data of all atoll sources with $N_H < 3\times 10^{22}$
cm$^{-2}$. It is clear that two distinct tracks have emerged, the
red track occurring on the shows a vertical transitions highlighting
that only the higher energies appear to be softening, whilst a
diagonal track is highlighted in blue, suggesting the spectrum is
softening at all energies, with the millisecond pulsars in green
overlaid on a grey back drop of all data. It should be noted that an
individual source will only follow one of these tracks.}
\label{fig:landr}
\end{figure*}

We identify and exclude data contaminated by dips and X-ray bursts by
using the standard 1 lightcurves (0.125s time resolution)
corresponding to each standard 2 spectra. We used the intrinsic
r.m.s. variability as a tracer of these. All lightcurves with r.m.s
above 50 percent were inspected for bursts, while the 16-s rebinned
lightcurves were checked for dips if their r.m.s. was above 20
percent.  Observations were also excluded if they were contaminated by
Galactic ridge emission, seen as a softening of the spectra and large
iron line contribution at the lowest luminosities in objects with low
Galactic latitude: see e.g. Wardzi{\'n}ski et al. (2002) for an
example of this in the black hole GX339-4.  Fig.  \ref{fig:landr}
shows the resulting colour-colour and colour-luminosity diagram for
all 34 atolls and 6 millisecond pulsars combined together. One
interesting thing to note is that several atoll sources approach or
even exceed Eddington luminosities (4U 1735-44; 4U 1744-26 (GX 3+1); 4U
1758-20 (GX 9+1); XTE J1806-246: Gladstone 2006), showing that
the high luminosity sources are not necessarily Z sources
(Hasinger \& van der Klis 1989).

Fig. \ref{fig:landr}(a) shows that while the atolls show the same
overall behaviour as claimed by DG03, they also show subtle but
significant differences. This is most evident in terms of the spectral
evolution during the transition between the hard (island) and soft
(banana) spectral states, which forms the middle branch of the total
Z-shaped track on the colour-colour diagram (even though these are
atolls, not Z sources, see Gierli{\'n}ski and Done 2002; Muno,
Remillard \& Chakrabarty 2002). We have highlighted this by picking out
in red those objects where the transition makes an almost vertical
track on the colour-colour diagram (hereafter called verticals), while
the blue symbols show those where the transition from island to banana
starts much further to the right on the colour-colour diagram, but ends
at approximately the same place, making a much more diagonal middle
branch (hereafter called diagonals). This difference in spectral
evolution is also picked out in the colour-luminosity diagram as a
different luminosity for the hard-soft transition. While the distances
are generally rather uncertain, this correlation between the behaviour
on the colour-colour and colour-luminosity diagrams give some
confidence in the reality of the luminosity difference. We stress that
this is not primarily due to hysteresis, the well known effect in
both black holes and neutron stars where the luminosity at which the
hard-soft transition occurs varies considerably in the same
object (e.g. Nowak 1995, Maccarone \& Coppi 2003). Here, the
transition luminosity is varying between different objects.

\begin{figure*}
\leavevmode
\begin{center}
\epsfxsize=10cm \epsfbox{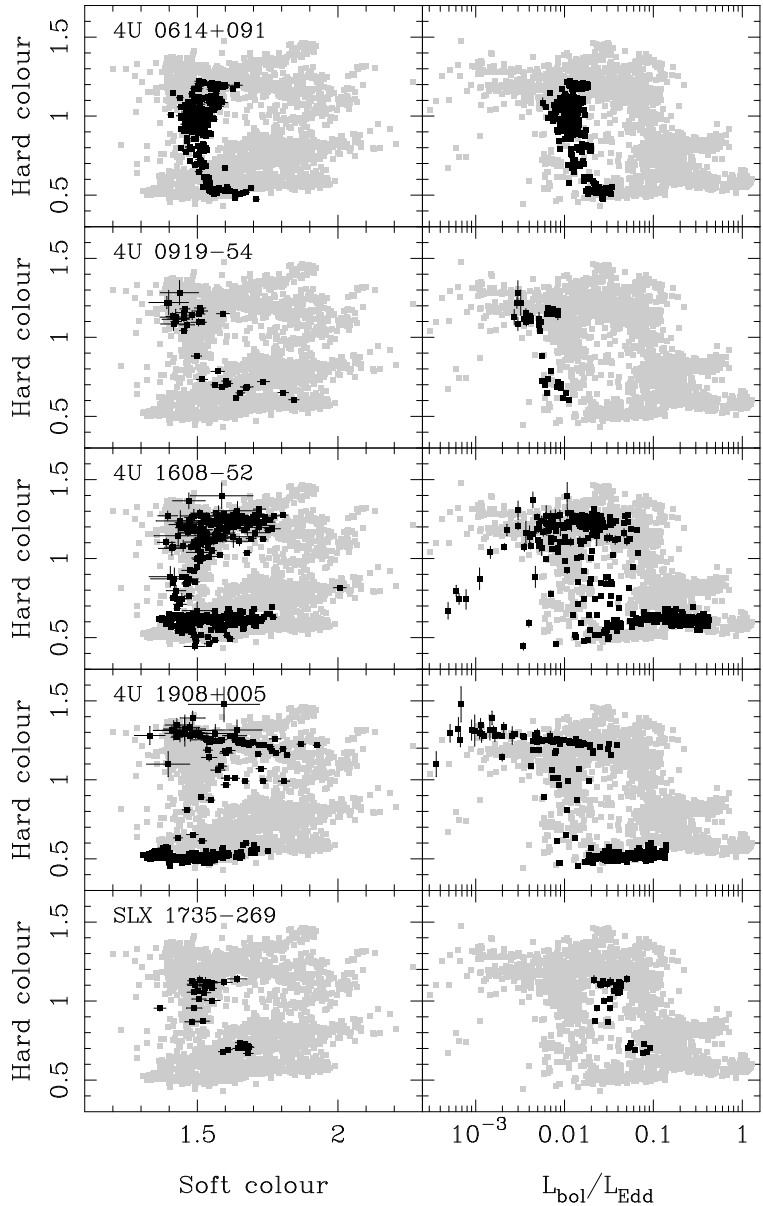}
\end{center}
\caption{Combined colour-colour colour-luminosity plots for the {\it
RXTE} PCA data of all atoll sources with $N_H < 3\times 10^{22}$
cm$^{-2}$ which display a vertical transition from hard to soft state
in the colour-colour diagram, plotted against a backdrop of all data.}
\label{fig:verticals}
\end{figure*}

\begin{figure*}
\leavevmode
\begin{center}
\epsfxsize=10cm \epsfbox{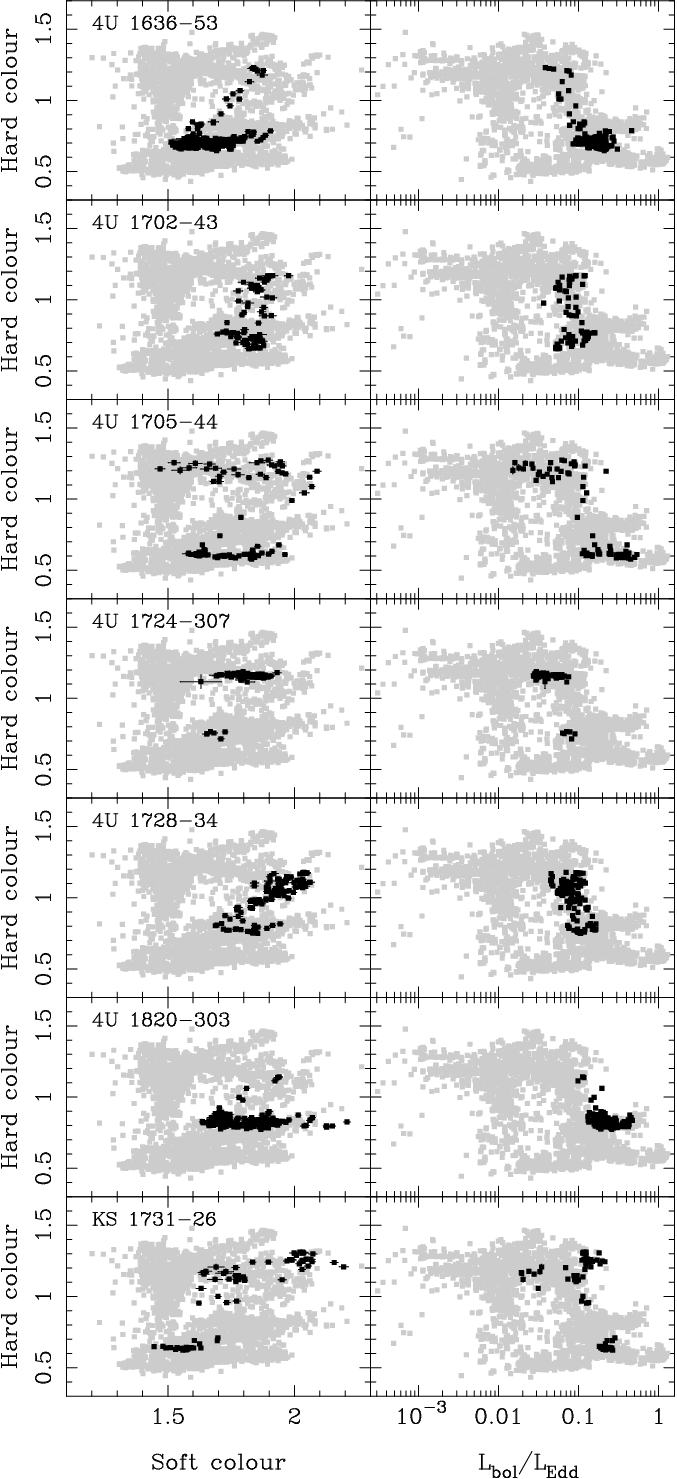}
\end{center}
\caption{Combined colour-colour colour-luminosity plots for the {\it
RXTE} PCA data of all atoll sources with $N_H < 3\times 10^{22}$
cm$^{-2}$  which display a diagonal transition from hard to soft
state in the colour-colour diagram, plotted against a
backdrop of all data.} \label{fig:diagonals}
\end{figure*}

\begin{figure*}
\leavevmode
\begin{center}
\epsfxsize=10cm \epsfbox{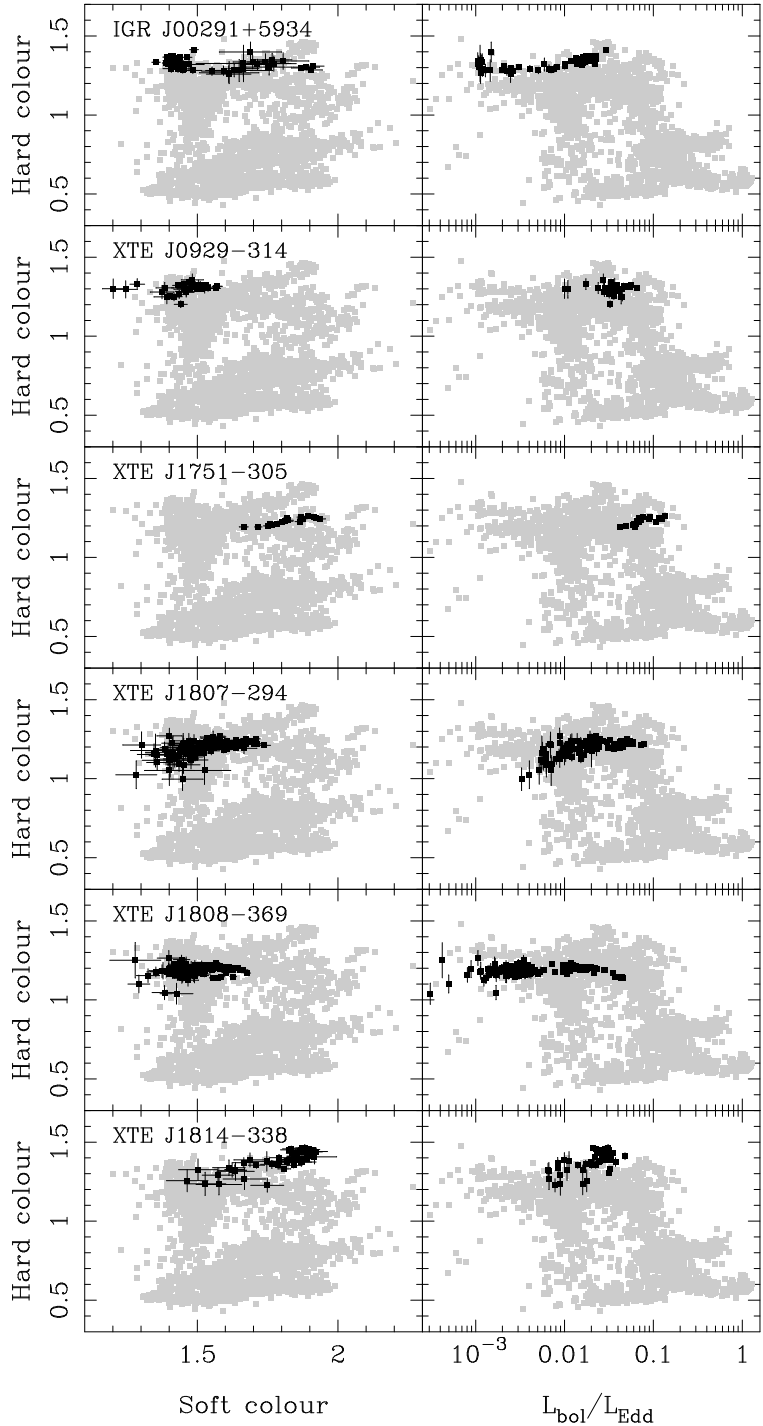}
\end{center}
\caption{Combined colour-colour colour-luminosity plots for the {\it
RXTE} PCA data of millisecond pulsars, plotted against a backdrop of
all data.} \label{fig:msps}
\end{figure*}

We show this difference in transition behaviour by plotting the
colour-colour and colour-luminosity diagrams for each individual
source which shows a clear state transition. Those making a vertical
transition are shown in Fig. \ref{fig:verticals}, while those making a
diagonal transition are shown in Fig. \ref{fig:diagonals}. Not all
objects have data covering a transition. Some, such as all the
millisecond pulsars, are only seen in the hard (island) state (see
Fig. \ref{fig:msps}), others e.g. 4U~1758-20 are seen only in the soft
(banana) branch. Of the ones which do make transitions, showing both
hard and soft spectra, the effects of data windowing mean that not all
have observations covering the transition period. Nonetheless, the
size of the {\it RXTE} database mean there are 12 sources which do
have enough data to constrain the shape of the transition on the
colour-colour diagrams.  The remaining sources, where there is
insufficient transition data available at present are included in the
background points plotted on each image, but not plotted individually here 
(they can be seen in Gladstone 2006).

\begin{figure}
\leavevmode
\begin{center}
\epsfxsize=8cm \epsfbox{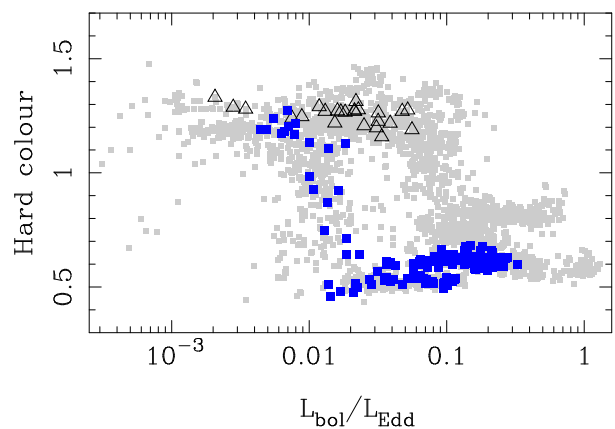}
\end{center}
\caption{Colour luminosity diagram highlighting the 
differences in behaviour of outburst rise and decay (hysteresis) in 
the bright transients 4U 1608--522 and 4U 1908+005.
The colour evolution during the fast rise at the onset of the outburst
(black triangles) matches that of the millisecond pulsars, while the
outburst decay (blue squares) matches the behaviour of the verticals.}
\label{fig:hysteresis}
\end{figure}

\subsection{Hysteresis}

The majority of sources have well defined transition luminosities
(to within a factor 2--3), irrespective of whether the transition is
from hard to soft i.e. on the rising part of the light curve, or
from soft to hard i.e. with a decreasing flux. However, there are
two sources where the transitions have a much large scatter in their
properties, namely 4U~1608--52 and 4U~1908+005 (Aql X-1). The colour-colour
and colour-luminosity diagrams for these are plotted as the bottom
two panels in Fig. \ref{fig:verticals}), but it is not clear from
the data that these sources should be classed as verticals, so here
we examine them in more detail.

These two systems are also the only two (apart from the millisecond
pulsars) which show large scale outbursts. In order to investigate
their scatter in transition behaviour we select simple outbursts from
the light curve, where there is a clear fast rise followed by a
monotonic decay. From this data we build a new, simplified
colour-luminosity diagram (Fig. \ref{fig:hysteresis}). The black
triangular points show the island state seen on the rise, which looks
very similar to the msps (Fig. \ref{fig:msps}). We then exclude all of
the transition data on the rise/peak of each outburst, and plot the
remaining banana branch and transition back to island state on the
decay as the blue square points.  Plainly the transition luminosity is
much larger in the rise to outburst than during the decay. This effect
of hysteresis is also seen in most Galactic black holes (e.g.
Maccarone \& Coppi 2003). Equally plainly, the decay shows a clear
vertical track, confirming the identification of these as verticals.
Hence we speculate that the msps are similar objects, and that if 
the msp outburst ever reached high enough luminosities to make a
transition then this transition would display hysteresis and 
the colour-colour track during the outburst decay would be vertical. 

None of the diagonals show marked hysteresis, though 4U 1705--44 and
KS 1731--26 may show a small effect in that their top branch (island
state) extends to higher luminosities than seen during the
transition (Fig. \ref{fig:diagonals}).  A similar small hysteresis
effect may also be present in the vertical 4U 0919--54.

\section{Discussion}

There are subtle differences in the spectral evolution of the atolls
on the colour-colour and colour-luminosity diagrams: there is the
distinction between the transition track (verticals and diagonals),
and then within the diagonals there is a difference in spectral
hardness of the banana branch, and within the verticals there are
objects which show large scale hysteresis. Here we examine possible
explanations for these effects.

\subsection{Inclination}

A range of inclinations is expected for these sources, and this could
be important for the observed spectrum if the intrinsic emission is
not isotropic.  Some degree of anisotropy is certainly expected from
the accretion disc due to its planar geometry, and as long as the
boundary layer has a different angular dependence then the overall
spectrum will change as a function of inclination.

To estimate the effects of inclination we use the spectral model
consisting of the disc emission ({\sc diskbb}) and the optically thick
boundary layer emission, which we model by thermal Comptonization ({\sc
comptt}). DG03 showed that the banana branch could be roughly
characterized by a disc varying in temperature in the range
1.0--1.5~keV together with an equal luminosity boundary layer with
$kT_e$ = 3 keV and $\tau$ = 5 (see also Revnivtsev \& Gilfanov (2006)). 
We use these parameters to model the shape of the banana branch.

\begin{figure*}
\leavevmode
\begin{center}
\epsfxsize=14cm \epsfbox{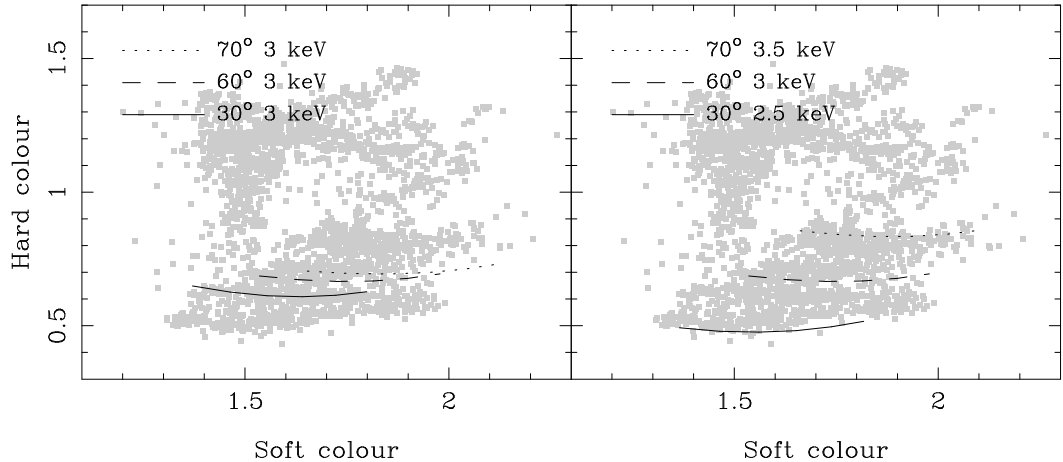}
\end{center}
\caption{Colour-colour diagrams showing the effect of inclination on
the position of the banana branch. The inclination is 30$^\circ$ (solid
curves), 60$^\circ$ (dashed curves) and 70$^\circ$ (dotted curves).
Each curve represents the disc plus boundary layer model, for the range
of disc temperatures of 1.0--1.5 keV (see text for details). In panel
(a) the disc emission is assumed to be angle-dependent, while the
boundary layer emission is isotropic, with constant temperature of 3
keV. In panel (b) the boundary layer emission is assumed to be
anisotropic, with higher observed temperature closer to the equator.
The three curves correspond to the same inclination angles as in panel
(a), but this time the boundary layer temperature is 2.5, 3.0 and 3.5
keV, for increasing angles, respectively.} \label{fig:inclination}
\end{figure*}

We first assume that the boundary layer is isotropic, while the disc
normalization varies as $\cos i$ but with angle averaged luminosity
equal to that of the boundary layer. Thus the apparent ratio of
disc to boundary layer flux will change as the inclination angle
varies. In Fig. \ref{fig:inclination}(a) we show how the banana branch
moves on the colour-colour diagram for inclinations of $30^\circ$,
$60^\circ$ and $70^\circ$. The change is mainly only in the soft
colour, with a greater proportion of disc emission shifting the start
of the banana branch to lower soft colours. Plainly this cannot account
for either of the two subtle spectral effects seen, but seems to have
more potential to explain the different positions of the banana branch
amongst the diagonals than the difference between verticals and
diagonals.

Examination of some of the individual banana branch spectra of the
diagonals showed that the Comptonization temperature of the boundary
layer is also changing, and in a way which is correlated with
$L_{\rm disc}/L_{\rm BL}$. This could indicate that the boundary layer
emission is itself anisotropic, perhaps with a temperature gradient so
that it is hotter close to the equatorial plane where the disc hits
the neutron star. Fig. \ref{fig:inclination}(b) shows the effect of
changing the temperature of the boundary layer from 2.5, 3.0 and
3.5~keV, for inclination angles of $30^\circ$, $60^\circ$ and
$70^\circ$, respectively. This matches very well with the range of
banana branches seen in the diagonals.

Testing this would require knowledge of the inclination, but very
few sources have good enough orbital determinations to constrain
this. However, any source exhibiting dips must be at fairly high
inclination ($i > 70^\circ$), and hence would be expected to have a
banana branch starting at fairly large hard and soft colours. This
can be seen in the case of 4U 1724--307 (see fig
\ref{fig:diagonals}), a dipping source whose banana branch ranges
from 1.65--1.75 in soft colour and 0.7 to 0.8 in hard colour. There
are three other dipping sources, 4U 1704--30, 4U 1746--37 and 4U
1915--05 (not included in this work), that also support this idea,
with each exhibiting large values of both hard and soft colours in 
the banana state (for details, see Gladstone 2006).

\subsection{Spin}

Burst oscillations and kHz QPOs have finally given observational
constraints on LMXB neutron star spins (e.g. the review by van der Klis
2000). For our sample, the inferred range is from 330 to 619~Hz.
However, this is unlikely to be the origin of the difference in
transition track as the spins seem fairly evenly distributed between
the two classes. However, spin could also give a difference in the
overall luminosity/temperature of the boundary layer. Lower spin gives
a higher relative speed between the inner edge of the disc and the
surface, leading to a lower $L_{\rm disc}/L_{\rm BL}$ ratio and a
higher boundary layer temperature, as required for the different banana
branch hard colours seen.  Thus it seems likely that some combination
of spin and inclination effects are responsible for the different
banana branches seen in the diagonals, but neither is likely to explain
the origin of the two different types of transition behaviour.

\subsection{Transient Behaviour and Hysteresis}

Of the subsample of systems with data covering the transition, the
publically available ASM and/or PCA Galactic Bulge long term light curves 
show large scale transient outbursts only in 4U 1608--52, 4U 1908+005 and
all the millisecond pulsars (see Table 1). This is very intriguing as
4U 1608--52 and 4U 1908+005 are both verticals. However, examination
of the long term lightcurves of the other verticals clearly shows that these
objects do not undergo dramatic outbursts, so this cannot be the
origin of the vertical/diagonal distinction.

However, 4U 1608--52 and 4U 1908+005 are also the only objects to show
clear large-scale hysteresis. Thus it seems possible that hysteresis
is linked to large amplitude outbursts.  There are no counterexamples
i.e. no systems that show major disc outbursts (which push the
accretion rate high enough to make a spectral transition) which do not
show hysteresis. All the millisecond pulsar outbursts remain in the
hard state, and all other atoll sources with enough data to deliniate
the transition and hence constrain hysteresis do not exhibit major
outbursts.

This hypothesis can be tested on the black hole LMXB systems.  Unlike
the atolls, none of these sources are persistant (McClintock \&
Remillard 2006).  This distinction in disc instability behaviour
between black holes and neutron stars can be broadly explained in the
context of the hydrogen ionization models.  Neutron stars are lower in
mass than black holes, so for the same companion star to overflow its
Roche lobe requires a smaller binary orbit for a neutron star compared
to a black hole. Smaller orbital separation means a smaller disc due
to tidal truncation. The size of the disc, together with the mass loss
rate from the companion star (determined by its evolutionary state)
determines the temperature of the coolest part of the disc. Thus the
smaller neutron star systems are less likely to have a disc which can
drop below the hydrogen ionization temperature required to trigger the
disc instability (King \& Ritter 1998; King, Kolb \& Szuszkiewicz
1997; Dubus et al 1999).

It is well known that most black hole systems show large scale
hysteresis (e.g. Maccarone \& Coppi 2003). Thus they support the idea
above that hysteresis is linked to large amplitude outbursts.  Even
more support comes from the one obvious exception, which is Cyg
X-1. This does not show hysteresis during its hard/soft transitions,
and is of course a persistent (HMXRB) source (e.g. Maccarone \& Coppi
2003; DG03).

Maccarone \& Coppi (2003) also suggested that hysteresis could be
linked to large scale outbursts, but with only Cyg X-1 as an
counterexample, the connection to outbursts is not quite so
unambiguous.  Cyg X-1 is also one of the few known Galactic black
holes with a high mass companion, so there is the possibility that
its accretion structure is somewhat different due to lower angular
momentum material from a stellar wind. By contrast, with the neutron
stars being generally stable to the hydrogen ionization trigger for
the disc outbursts, there are many counterexamples, all with low
mass companions. We can also rule out the alternative origin for
hysteresis discussed by Maccarone \& Coppi (2003), namely that it is
produced simply by a difference in behaviour in low mass X-ray
binaries between a hard--to--soft transition on the rising part of
the light curve, and a soft--to--hard transition as the flux
decreases. Most of the sources shown in Figs. \ref{fig:diagonals}
and \ref{fig:verticals} make the transition in both directions, and
do not show large scale hysteresis (though there are smaller scale
effects, consistent with the difference in environment: Meyer \&
Meyer-Hoffmeister 2005). Instead, hysteresis is seen where the mass
accretion rate changes dramatically (from $\sim L/L_{\rm Edd}\ll
10^{-4}$ to $\ga0.1$). We speculate that the accretion flow is able
to access non-equilibrium states during the rapid changes in
accretion disc structure caused by the disc instability.

Thus the neutron stars clearly show a one--to--one correlation between
the dramatic large amplitude outbursts triggered by the disc
instability and hysteresis, consistent this being the origin of
hysteresis. This predicts that the millisecond pulsar outbursts should
also show hysteresis if their accretion rate at the outburst peak ever
goes high enough to trigger a spectral transition (in which case we
would expect them also be verticals, along with 4U 1608--52 and 4U
1908+005). However, there is no such one--to--one correlation between
outbursts and transition behaviour (vertical/diagonal), so
we explore further aspects of the systems below.

\subsection{Binary system parameters}

Table \ref{tab:sources} shows the binary parameters where these are
known. We first explore whether there is any correlation between
transition behaviour and binary period (which is a tracer of
companion type, and also of superburst behaviour: Cumming 2003). For
the verticals, the periods span a large range, from 1 hour for the
persistent sources 4U 0614+091 and 4U 0919--54 through to 19 and 288
hours for the transients 4U 1608--52 and 4U 1908+005. This
distinction is as expected from the disc instability model (wide
binary implies a much larger disc, so a cooler outer edge which is
more likely to trigger the hydrogen instability). If the millisecond
pulsars are also verticals then these fill in the period
distribution from 40 minutes to 4 hours.There is very little data to
compare this to the diagonals, as only 2 have periods (11 minutes
and 3.9 hours), but the broad span seen from the verticals
encompasses most of the binary periods seen in atolls, so binary
period alone is unlikely to be the origin of the transition track
dichotomy.

\subsection{Long-term mass accretion rate}

The distinction between millisecond pulsars, which plainly retain a
residual magnetic field channelling the flow, and other atolls which
do not show pulsations is most probably due to the long timescale
mass accretion rate (Cumming, Zweibel \& Bildsten 2001). Plausibly,
high accretion rates can bury the magnetic field below the neutron
star surface, but the very low average mass accretion rate in the
millisecond pulsars is insufficient to bury the field (Cumming et
al. 2001). Thus there is a potential physical mechanism for the long
term mass accretion rate to change the properties of the accretion
flow.

We estimate the long-term mass accretion rate, $\langle \dot{m}
\rangle$, (or corresponding average luminosity, $\langle L_{\rm
bol}/L_{\rm Edd}\rangle$) for all of the systems from the observed
X-ray emission. The PCA data gives estimates for $L_{\rm bol}$
through spectral fitting, but the light curves are highly
incomplete, so the data cannot simply be used as an indicator of the
long term average mass accretion rate. By contrast, the {\it RXTE}
All Sky Monitor (ASM) gives an almost continuous light curve for
every bright X-ray source in its field of view, but its lack of
spectral resolution means that going from count rate to $L_{\rm
bol}$ is highly uncertain. Hence we combine the two approaches. We
use the PCA light curve to define the average $\langle L_{\rm
bol}/L_{\rm Edd}\rangle$ during these observations, then select the
simultaneous ASM points to find the average ASM count rate during
the PCA observations. The ratio of this to the full ASM light curve
gives the correction for the incompleteness of the PCA observations.

This approach works well unless the source becomes very faint, in
which case contamination of the ASM by other nearby sources and/or
galactic ridge emission can be a problem. The only sources for which
this is an issue are the transients i.e. the verticals 4U 1608--52
and 4U 1908+005 and all the millisecond pulsars. The outbursts of 4U
1608--52 and 4U 1908+005 are so bright that they dominate the
average of the ASM lightcurves. However, this is not the case for
the millisecond pulsars, so for these we use a different approach.
Here we use the PCA data alone, which have good outburst coverage,
assuming that the source intensity is negligible in the periods
outside the known outbursts. The results of both the millisecond
pulsars and atoll sources are shown in Table \ref{tab:lbol}.

\begin{table}
\leavevmode
\begin{minipage}{85 mm}
\begin{center}
\begin{tabular}[width=\textwidth]{ccc}
\hline
Source Name & Source Type & $\langle L_{\rm bol}/L_{\rm Edd}\rangle$ \\
\hline
\\
IGR J00291+5934    & msp & 1.4 $\times 10^{-5}$ \\
XTE J0929--314     & msp & 1.4 $\times 10^{-4}$ \\
XTE J1751--305     & msp & 5.2 $\times 10^{-5}$ \\
XTE J1807--294     & msp & 2.7 $\times 10^{-4}$ \\
XTE J1808--369     & msp & 6.7 $\times 10^{-5}$ \\
XTE J1814--338     & msp & 1.6 $\times 10^{-4}$ \\
\\
4U 0614+091 & V & 3.2 $\times 10^{-3}$ \\
4U 0919--54 & V & 3.1 $\times 10^{-3}$ \\
4U 1608--52 & V & 6.6 $\times 10^{-3}$ \\
4U 1908+005 & V & 3.7 $\times 10^{-3}$ \\
SLX 1735-269 & V & 1.6 $\times 10^{-3}$ \\
\\
4U 1636--53  & D & 4.4 $\times 10^{-2}$ \\
4U 1702--43  & D & 7.8 $\times 10^{-3}$ \\
4U 1705--44  & D & 1.9 $\times 10^{-2}$ \\
4U 1724--307 & D & 7.3 $\times 10^{-3}$ \\
4U 1728--34  & D & 1.9 $\times 10^{-2}$ \\
4U 1820--303 & D & 1.3 $\times 10^{-1}$ \\
KS 1731--26  & D & 3.4 $\times 10^{-2}$ \\
\\

\end{tabular}
\end{center}
\caption{Long-term average luminosity revealing a distinct break
between millisecond pulsars (msp) and atoll sources as well as a
more subtle break between sources where the hard--soft transition is
approximately vertical on a colour-colour diagram (V) and those
where it makes a diagonal track (D).} \label{tab:lbol}
\end{minipage}
\end{table}

It can be seen that there is a systematic difference between the
three classes of sources. The millisecond pulsars have the lowest
$\langle L_{\rm bol}/L_{\rm Edd}\rangle$, then the verticals, then
the diagonals.  Thus it seems possible that this is the origin of
the difference in transition properties. The millisecond pulsars'
low $\langle \dot{m} \rangle$ allows the field to diffuse out of the
crust, and to be strong enough to collimate the flow and produce
pulsations. One possibility might be that while the high $\langle
\dot{m} \rangle$ of the diagonals suppresses the field entirely, the
medium $\langle \dot{m} \rangle$ in the verticals allows some field
to diffuse out. But the verticals show no trace of pulsation, so the
magnetic field cannot collimate any significant part of the flow.

One way around these pulsation limits is to separate the field and
accretion flow. Any non-spherical accretion flow will predominantly
bury the field in the region of the flow, leaving the field to
escape in regions with little accretion. Even the hot accretion flow
envisaged for the hard island states favours the equatorial plane,
with a geometry which is more like a thick disc than a truly
spherical flow (e.g. Narayan \& Yi 1995). This nicely circumvents
the pulsation limits, but then there is no interaction between the
magnetic field and the flow, and so no physical mechanism to change
the behaviour of the transition.

However, one aspect of the system that a polar magnetic field could
affect is the jet. The most recent numerical simulations of the
accretion flow magnetohydrodynamics show a causal link between the
jet and accretion flow (Hawley \& Krolik 2006; McKinney 2006), so
this might give an indirect link between the magnetic field and
accretion flow properties.  One way to test this is to look at the
radio emission from these systems. While theoretical models of jets
are not well developed, we can use the observed GBH behaviour as a
template. These show a clear correlation between radio and X-ray
luminosity in their low/hard state, showing that $L_X/L_{\rm Edd}$
is important in determining the power of the jet (Gallo et al.
2006). Hence to see whether there are differences in the jet
emission between the millisecond pulsars, verticals and diagonals we
need to compare the radio emission at the same $L_X/L_{\rm Edd}$.
Hysteresis gives potential problems, so ideally the comparison would
be between persistent verticals and diagonals in the hard island
state at the same $L_X$. However, there is very little data to make
this comparison, with only 4U 0614+091 and 4U 1728--34 for the
persistent verticals and diagonals, respectively, and these do not
overlap in $L_X$ (Migliari \& Fender 2006). Thus we speculate that
the origin of the difference in transition behaviour between the
verticals and diagonals is due to the presence of some magnetic
field at the pole in the verticals which affects the accretion flow
indirectly through jet formation, by contrast to the diagonals which
have higher mass accretion rates, sufficient to bury the field
everywhere.

\section{Conclusions}

The atolls and millisecond pulsars are consistent with showing the
same overall spectral evolution with changing mass accretion rate,
but there are some subtle differences. The spectral shape of the
soft (banana) branch shows subtle variations from object to object,
most probably due to combination of inclination and spin changing
the ratio of the observed disc to boundary layer luminosity, and
boundary layer temperature. However, there are also clear
differences in behaviour during the hard/soft transition which point
to a more fundamental distinction. The data shows two types of
sources, those where the transition makes a vertical track on the
colour-colour diagram, and occurs at $L/L_{\rm Edd}\sim 0.02$, and
those which make a diagonal track on the colour-colour diagram with
the transition at $L/L_{\rm Edd}\sim 0.1$. There are hysteresis
effects in individual sources which introduce dispersion in the
transition luminosity, but these are large {\em only} for the
outbursting atolls (which are both verticals). Splitting these
outbursts into the rapid rise and slow decay phases show that the
rapid rise looks like the millisecond pulsars (so they are probably
also verticals) while the slow decay looks like the persistent
verticals.  Thus it seems likely that large scale hysteresis effects
are only seen in sources where the disc structure changes rapidly
due to the onset of the hydrogen ionization instability. This is
also consistent with the observed black hole behaviour. The
association of the millisecond pulsars with verticals suggests that
the difference in transition is ultimately linked to the surface
magnetic field, and indeed, all the verticals have long term mass
accretion rates which are smaller than those of the diagonals,
though not as small as those of the millisecond pulsars. Thus the
verticals could have some small B field which is able to affect the
inner accretion flow, but this must be indirect as otherwise these
systems also would show pulsations. We speculate that the physical
link between the magnetic field (predominantly polar) and accretion
flow (predominantly equatorial) may be due to the changes in the
jet, which would be testable with more radio data on these sources.

\section*{Acknowledgements}

This research has made use of data obtained through the High Energy
Astrophysics Science Archive Research Center Online Service, provided
by the NASA/Goddard Space Flight Center.


\label{lastpage}

\end{document}